\documentclass[10pt, conference, letterpaper]{IEEEtran}

\usepackage{geometry}
\usepackage{epsfig}
\usepackage{amssymb,amsmath,amsfonts}
\usepackage{amsthm}
\usepackage{mathrsfs}
\usepackage{graphicx}
\usepackage{multirow}
\usepackage{color}
\usepackage{adjustbox}
\usepackage{enumitem}

\usepackage{epsfig} 
\usepackage{mathrsfs}
\usepackage{graphicx}
\usepackage{multirow}
\usepackage{mathtools}
\usepackage{caption}
\usepackage{subcaption}
\usepackage{booktabs}
\usepackage{caption}
\usepackage[font=footnotesize]{caption}

\usepackage{framed}
\usepackage{pbox}
\usepackage{tikz}
\usepackage{graphicx,adjustbox}
\usetikzlibrary{decorations.pathreplacing}
\usepackage{float}
\usepackage{anyfontsize}
\usepackage{hyperref}
\usepackage[
ruled,vlined]{algorithm2e}

\newcommand{\mg}{\color{magenta}}
\newcommand{\rd}{\color{red}}
\newcommand{\bl}{\color{blue}}

\usepackage{dsfont}
\usepackage{algorithmicx}
\usepackage{soul}
\newcommand{\pr}[1]{\mathrm{Pr}#1} 
\newcommand{\vect}[1]{\boldsymbol{\mathrm{#1}}}
\newcommand{\mc}[1]{\mathcal{#1}}
\newcommand{\ms}[1]{\mathsf{#1}}
\newcommand{\mb}[1]{\mathbf{#1}}

\newcommand{\sd}{\mathbf{SD}}
\DeclareMathAlphabet{\mathpzc}{OT1}{pzc}{m}{it}
\newcommand{\kem}{\mathsf{K}}
\newcommand{\ekem}{\mathsf{K.Enc}}
\newcommand{\dkem}{\mathsf{K.Dec}}
\newcommand{\gkem}{\mathsf{K.Gen}}
\newcommand{\lkem}{\mathsf{K.Len}}

\newcommand{\ikem}{\mathsf{iK}}
\newcommand{\ikemg}{\mathsf{iK.Gen}}
\newcommand{\ikeme}{\mathsf{iK.Enc}}
\newcommand{\ikemd}{\mathsf{iK.Dec}}
\newcommand{\likem}{\mathsf{iK.Len}}

\newcommand{\ikemgo}{\mathsf{iK_{OWSKA}.Gen}}
\newcommand{\ikemeo}{\mathsf{iK_{OWSKA}.Enc}}
\newcommand{\ikemdo}{\mathsf{iK_{OWSKA}.Dec}}

\newcommand{\dem}{\mathsf{D}}
\newcommand{\ddem}{\mathsf{D.Dec}}
\newcommand{\edem}{\mathsf{D.Enc}}
\newcommand{\gdem}{\mathsf{D.Gen}}
\newcommand{\ldem}{\mathsf{D.Len}}

\newcommand{\ike}{\mathsf{HE}}
\newcommand{\gike}{\mathsf{HE.Gen}}
\newcommand{\eike}{\mathsf{HE.Enc}}
\newcommand{\dike}{\mathsf{HE.Dec}}

\newcommand{\hpk}{\mathsf{HPKE}}
\newcommand{\ghpk}{\mathsf{HPKE.Gen}}
\newcommand{\ehpk}{\mathsf{HPKE.Enc}}
\newcommand{\dhpk}{\mathsf{HPKE.Dec}}

\newcommand{\remove}[1]{}

\newcommand{\eps}{\epsilon}

\providecommand{\eqref}[1]{(\ref{#1})}
\newtheorem{construction}{Construction}

\newcommand{\advu}{\mathsf{A^U}}
\newcommand{\advbuone}{\mathsf{B^{U}_1}}
\newcommand{\advbutwo}{\mathsf{B^{U}_2}}
\newcommand{\advbu}{\mathsf{B^{U}}}

\newcommand{\advuone}{\mathsf{A^U_1}}
\newcommand{\advutwo}{\mathsf{A^U_2}}
\newcommand{\sadvuone}{[\mathsf{A^U_1}]}
\newcommand{\sadvutwo}{[\mathsf{A^U_2}]}
\newcommand{\msa}{\mathsf{A}}
\newcommand{\advb}{\mathsf{A^B}}
\newcommand{\advbone}{\mathsf{A^B_1}}
\newcommand{\advbtwo}{\mathsf{A^B_2}}
\newcommand{\sadvbone}{[\mathsf{A^B_1}]}
\newcommand{\sadvbtwo}{[\mathsf{A^B_2}]}
\newtheorem{theorem}{Theorem}
\newtheorem{definition}{Definition}
\newtheorem{lemma}{Lemma}
\newtheorem{remark}{Remark}
\newtheorem{proposition}{Proposition}
\newtheorem{corol}{Corollary}

\begin{document}
\title{Information-theoretic Key Encapsulation and  its Applications} 

\author{Setareh Sharifian, Reihaneh Safavi-Naini
  \IEEEauthorblockN{}
  \IEEEauthorblockA{University of Calgary, Canada}
}


\maketitle

%
%




\begin{abstract}
A {\em  hybrid encryption scheme}   is a public-key encryption system that consists of a public-key part   called the {\em key encapsulation mechanism (KEM)}, 
and a (symmetric) secret-key part called  {\em data encapsulation mechanism (DEM)}:  the public-key part is used to generate a shared secret key between two parties, and the symmetric key part is used to encrypt the message using the generated key.   Hybrid encryption schemes are widely used for secure communication over the Internet. 
In this paper, we initiate the study of {\em hybrid encryption  in preprocessing model} which assumes access to initial correlated variables by all  parties (including the eavesdropper). 
We define {information-theoretic KEM (iKEM)} that, together with a (computationally) secure DEM, results in a  hybrid encryption scheme in preprocessing model. 
We define the security of each building block, and prove a
composition theorem that guarantees (computational) $q_e$-chosen-plaintext  (CPA) security of the hybrid 
encryption system if the iKEM and the DEM satisfy  $q_e$-chosen-encapculation attack and one-time security, respectively.
We show that iKEM can be realized by a one-way SKA (OW-SKA) protocol 
with a revised security definition. Using an OW-SKA that satisfies this revised definition of security effectively allows the secret key that is generated by the OW-SKA to be used with a one-time symmetric key encryption system such as XORing a pseudorandom string with the message, and provide $q_e$-CPA security for the hybrid encryption system. We discuss our results and directions for future work.
\end{abstract}

\section{Introduction}

Public-key encryption (PKE) schemes  are usually defined 
for  restricted message spaces and so the 
ciphertext can hide a limited number of plaintext bits.
A {\em  Hybrid encryption scheme}   consists of a public-key part 
and a (symmetric) secret-key part. 
The public-key part is called   {\em key encapsulation mechanism (KEM)}  and generates a pair of  (i)  a random symmetric key $K$, and  (ii) a ciphertext $c$.
The symmetric key part uses the generated key $K$ to encrypt the actual data  and obtains the corresponding ciphertext $c'$ using an efficient {\em data encapsulation mechanism (DEM)} (e.g. that can be constructed  as counter mode of AES \cite{standard2001announcing}).  The pair $(c,c')$ allows the decryptor to first recover $K$ from $c$, and then use it to decrypt $c'$ and obtain the data.
{\em KEM/DEM paradigm} was formalized by Cramer and Shoup \cite{cramer2003design} and has been widely used  in Internet protocols to implement public-key encryotion in protocols such as   TLS   \cite{tls} and SSH protocols \cite{ssh}, and is
  incorporated in   standards such as \cite{irtf-cfrg-hpke-07}.


Today's main constructions of KEM  rely on the hardness assumption of  two computational problems, discrete logarithm (DL)  and integer factorization problems,  for  both of which efficient quantum algorithms have been given by Shor \cite{shor1999polynomial}.
KEMs that remain secure in presence of quantum computers (are  {\em post-quantum secure}) have  been constructed using hard problems in areas such as lattices or algebraic codes for which efficient
quantum algorithms are not known
 \cite{bernstein2016ntru,bos2018crystals,aragon2017bike,melchor2018hamming}.
 These constructions in many cases require high computation and communication cost \cite{khalid2019lattice} and in all cases need to update their parameters with advances in computing and security technologies (e.g. updating parameters
when new algorithms and attacks are found). 
We define  post-quantum security as security against an adversary with access to a quantum computer,  and note that post-quantum security of  hybrid encryption systems is primarily determined by the post-quantum security of the  KEM because DEM is constructed using a symmetric key  encryption that its security will not be significantly affected by quantum computers (one needs to increase the key size).

In this paper, {\em we initiate the study of  ``KEM/DEM paradigm in preprocessing model''}, where the specification of the protocol includes a joint distribution $\mc P$ over $R_1\times..\times R_n$, where $R_i,\ i=1\dots, n$, is a finite domain. 
A (trusted) sampler samples correlated random values $(r_1,\dots, r_n)$, and delivers $r_i$ to the party $P_i$ before the protocol starts (thus making it independent of the input).  
 The   model has been widely studied in cryptography with both positive and negative results on unconditionally secure computation with correlated randomness \cite{ishai2013power}. 
 Source model  in information theoretic key agreement \cite{Ahlswede1993,Maurer1993} uses a similar initial setup.
 In a two party key agreement in source model,  before protocol starts, 
 a trusted sampler samples a public distribution $P_{XYZ}$ and gives the samples of $x$, $y$, and $z$, to Alice, Bob and Eve, respectively.
 An example of this setting was considered by Maurer \cite{Maurer1993} and is known  as {\em satellite setting},  where a satellite broadcasts a random beacon that is received by protocol parties through their (independent) channels.
One way SKA (OW-SKA) \cite{Holenstein2006,holenstein2006strengthening} is a two-party key agreement that transmits a single message from Alice to Bob, and if can be used  as a KEM, together with DEM can provide a hybrid encryption scheme with post-quantum security. Intuitively,  using an information-theoretically secure KEM will establish a key whose security will not depend on the computational power of the adversary, and  since a DEM component that is implemented using an algorithm such as AES-256  will be 
 safe against quantum computers  \cite{bonnetain2019quantum},
  the combination of the two will provide post-quantum security.

{\em KEM with private input.} 
KEM in preprocessing model will use private randomness samples (correlated random variables) of Alice and Bob to establish a shared key and so \textit{hybrid encryption scheme in preprocessing model} will be neither a public-key, nor  a symmetric key system (which would require a shared secret key before the scheme is used) but will use correlated samples of the two parties as the initial setup.
The hybrid encryption system will be {\em computationally secure} (although it can be extended to information-theoretic case (see Discussion in Section~\ref{sec:conclude}).
 A traditional OW-SKA cannot be directly used as a KEM  because of the difference in the security definition of the two.

\vspace{0.5mm} 
\noindent
{\bf Overview of our main contributions.}  
We formalize KEM/KEM paradigm in preprocessing model and prove a composition theorem for $q_e$-CPA security of the hybrid encryption system  using an information-theoretic KEM (iKEM) with security against $q_e$ encapsulation queries and a one-time DEM.
Security notion of iKEM (Definitions~\ref{def:sikem}), similar to the computational KEM, uses  game-based security notion and defines security by bounding the success probability of an adversary in a game played against a challenger that is described as a probabilistic experiment with a well-defined success event.  Lemma~\ref{lem:sdeno} relates the success probability of the adversary to the statistical distance based security definition of OW-SKA.  
Security definition of DEM will be the same
as in traditional hybrid encryption schemes and is recalled in Definition~\ref{def:demsec}. We define the hybrid encryption in preprocessing model (Definiton~\ref{def:he}) and its security notions  (Definiton~\ref{def:she}) against a computationally bounded adversary, and prove a composition theorem (Theorem~\ref{theo:composition}) that shows combining    an iKEM and a DEM with appropriate security  definitions results in a   hybrid encryption system in preprocessing model with provable security with respect to the defined security notion.

In Section \ref{sec:const}, we construct an iKEM  by modifying the OW-SKA in \cite{sharif2020} to provide iKEM security in the sense of Definition~\ref{def:sikem}.  Parameters of the iKEM can be chosen to output a secure key of length, for example, 256 bit, to be used to construct a pseudorandom sequence that will be XORed with the message to provide post-quantum  $q_e$-CPA security. 

  \vspace{0.5mm} 
\noindent
{\bf Discussion.} 
Hybrid encryption scheme in preprocessing model allows
secure application of information-theoretic secure OW-SKA with symmetric
key encryption systems.  It had been noted \cite{maurer2011constructive,ben2005universal,renner2005universally} that, using a secret key that is established by an information-theoretic key agreement,  in an encryption algorithm,  requires stronger security definition for key agreement  to  guarantee security of encryption and allow composability of the key agreement protocol and the encryption system.   
Security definition of iKEM allows secure composition of a (OW-SKA)  information theoretic key  agreement with a computationally secure symmetric key encryption to achieve secure hybrid encryption. 

\textit{Extensions.}  A hybrid encryption scheme in preprocessing model can be seen as a private key encryption scheme that uses the private input of Alice  to encrypt a  message,  and the unequal but correlated private input of Bob, to decrypt the ciphertext.  
Maurer \cite{Maurer1993} showed that secret key agreement with information-theoretic security requires initial correlated variables. 
Our proposed construction of iKEM in source model can be realized in wireless settings using physical layer correlation generation setups.
iKEM can also be constructed using other setup assumptions such as the setup of fuzzy extractors \cite{dodis2004fuzzy}  where Alice and Bob have ``close'' variables of sufficient min-entropy. 
Our work can be extended to define stronger security notions (e.g. security against chosen ciphertext attack (CCA)) for iKEM and hybrid encryption in preprocessing model.

\textit{Applications.} Hybrid encryption in preprocessing model gives an efficient encryption scheme with post-quantum security that will be particularly attractive for resource constrained IoT
devices such as a smart lock with long life and so need for
postquantum security. 

\vspace{1mm} 
{\noindent 
\bf Related Works.} 
 Cramer and Shoup  \cite{cramer2003design} formalized KEM/DEM paradigm  and proved that CCA security of KEM and DEM  as a public-key and a symmetric key encryption systems, respectively, leads to CCA security of the final (public  key) hybrid  encryption system. This is  the strongest commonly used security notion for encryption system.
 There is a large body of work on construction of CCA-secure hybrid encryption using different security requirements for KEM  and DEM \cite{kurosawa2004new},\cite{herranz2006kurosawa,abe2005tag,shacham2007cramer}.
CPA security of hybrid encryption and its relation with other security notions of hybrid encryption is studied in \cite{herranz2006kem}.

Study of secret key agreement in source model was initiated by Maurer \cite{Maurer1993} and  independently by Ahlswede and Csisz{\'{a}}r \cite{Ahlswede1993},  with many followup works for different physical layer setups.
One-way secret key (OW-SK) capacity was defined and derived  by Ahlswede and Csisz{\'{a}}r \cite{Ahlswede1993}.
Holenstein and Renner  \cite{Holenstein2006} considered one-way SKA (OW-SKA) protocols 
and gave constructions that achieve  OW-SK capacity. 
There are a number of capacity achieving OW-SKA constructions  \cite{Renner2005,renes2013efficient,Chou2015a,sharif2020}, in some cases  \cite{sharif2020} with explicit lower bound on finite key length.

Cryptographic premitives that have been studied in preprocessing model, 
 include oblivious transfer \cite{beaver1995precomputing,rivest1999unconditionally} and multi-party computation (MPC) protocols \cite{bendlin2011semi,ishai2013power,garg2018two}. 
 The source model in information-theoretic key agreements uses  a similar 
  initialization phase 
  \cite{maurer1997information}.


\vspace{1mm} 
{\noindent 
\bf Organization.}
Preliminaries are reviewed in is given in Section~\ref{sec: pre}. 
In Section~\ref{sec:new},  we propose  hybrid encryption in preprocessing model and discuss its security.
 A 
 concrete  construction of  iKEM 
  is given in Section~\ref{sec:const},
and concluding remarks are discussed in  Section~\ref{sec:conclude}.

\section{Preliminaries}\label{sec: pre}

{\em Notations.} 
	 We denote random variables (RVs) with upper-case letters, (e.g., $X$), and their realizations with lower-case letters, (e.g., $x$). 
	 Calligraphic letters denote sets, 
	 and  size of a set $\mc X$ is denoted by $|\mc X|$. 
	 	 $U_{\mathcal{X}}$ and $U_\ell$ denote uniformly distributed  random variable  over ${\mathcal{X}}$ and  $\{0,1\}^\ell$, respectively. 
	 	Bold face letters denote vectors, and $\vect{X}=X^n=(X_1,\ldots,X_n)$  is a vector of  $n$ random variables, and its instantiation is given by $\vect{x}= x^n = (x_1,\ldots,x_n)$. 
	    The set $\mc X^n$ is the $n$ times Cartesian product of $\mc X$. 
	    An information source $\vect X$ is IID if for any $n>0$, the sequence of $n$  source outputs, $X_1,\ldots,X_n$, are independent and identically distributed. The sequence  is called $n-$IID sample. A function $\ms F:\mc X\to \mc Y$ maps an element $x\in \mc X$ to a corresponding element $y\in \mc Y$. This is denoted by $ y=\ms F(x)$. If $\ms A$ is a probabilistic algorithm, then $\ms A(x_1,x_2,\dots;r)$ is the result of running $\ms A$ on inputs $x_1,x_2,\dots$ and coins $r$. We say that $y$ can be output by $\ms A(x_1,x_2,\dots)$ if there is some $r$ such that $\ms A(x_1,x_2,...;r) = y$.
  We use the symbol  `$\leftarrow$', to
assign a constant value (on the right-hand side) to a variable (on the left-hand side). Similarly, 
we use, `$\stackrel{\$}\leftarrow$', to assign to a variable either a uniformly
sampled value from a set or the output of a randomized algorithm. 
We denote by $x\stackrel{r}\gets \mathrm{P}_X$ the assignment of a  
sample from $\mathrm{P}_X$ to the variable $x$.
We write $\msa^{\ms O_1,\ms O_2,...}(.)$ to denote an { algorithm} $\msa$  that has access to oracles $\ms O_1, \ms O_2, . . . $,
and by  $u
\gets\msa^{\ms O_1,\ms O_2,...}(x, y, \cdots)$ denoting the algorithm taking inputs $x, y, \cdots$, and generating output $u$. 
	 %


The probability mass function (p.m.f) of a random variable (RV) 
	   $X$ is denoted by $\mathrm{P}_X$ and $\mathrm{P}_X(x) = \pr(X=x)$. The p.m.f corresponding to $X^n$ with distribution $P_X$ is denoted by $P_X^n$.
	   
For two random variables $X$ and $Y$, $\mathrm{P}_{XY}$ denotes their joint distribution, and $\mathrm{P}_{X|Y}$ denotes their conditional distribution. 
 The {\em statistical distance} between  two 
  RVs $X$ and $Y$ that are defined over 
on a set 
$\mc T$, is given by, 
\begin{align}\label{eq:sd}
    \sd(X;Y) 
    =\displaystyle\max_{\mc W\subset\mc T}(\pr(X\in \mc W)-\pr(Y\in \mc W)), 
\end{align}
where $\pr(X\in \mc W)=\sum_{t\in \mc W}\pr(X=t)$.

\remove{
{\mg
The following is a corollary from \cite[Lemma 6.2]{cramer2003design} that will be used in our proofs.
\begin{lemma}\cite[Lemma 6.2]{cramer2003design}\label{lem:event}
Let $U_1$, $U_2$, and $F$ be random variables defined on some probability space. Suppose that the event $U_1\wedge \neg F$ occurs if and only if $U_2\wedge \neg F$ occurs. Then $|\pr[U_1]-\pr[U_2]|\leq \pr[F]$.
\end{lemma}
}}


We define  two classes, ${SMALL}$ and ${NEGL}$ of functions, and use  them to bound closeness of distribution and define statistical and computational indistinguishability, respectively. These definitions  follow  Definition
6.1 in \cite{pfitzmann2000model}.
A set ${SMALL}$ of functions $\mathbb{N}\to \mathbb{R}_{\geq 0}$ is a class of \textit{small functions} if it is closed under addition, and with a function
$\ms g$ also contains every function $\ms g'$ with $\ms g'\leq \ms g$.
The class ${NEGL}$ of negligible functions contains all functions $\ms f: \mathbb{N}\to \mathbb{R}_{\geq 0}$ that decrease faster than the inverse of every polynomial, i.e., for all positive polynomials $\ms P(\cdot)\ \exists \lambda_0\ \forall \lambda>\lambda_0: \ms f(\lambda)<\frac{1}{\ms p(\lambda)}$.

The  \emph{min-entropy} $H_{\infty}(X)$ of a random variable $X \in \mathcal{X}$ with distribution $\mathrm{P}_X$
is  defined by  
$H_{\infty}(X)= -\log (\displaystyle \max_{x} (\mathrm{P}_X({x})))$.
The \emph{average conditional min-entropy}  \cite{dodis2004fuzzy} is commonly defined as,
$\tilde{H}_{\infty}(X|Y)= -\log \displaystyle \mathbb{E}_{{y} \in \mathcal{Y}}\displaystyle \max_{{x} \in \mathcal{X}}\mathrm{P}_{X|Y}({x}|{y}).$
The following lemma, proven in \cite{dodis2004fuzzy}, gives useful properties of the average conditional min-entropy that will be used in our proofs.
 \begin{lemma}\label{lem:entropy}\cite[Lemma2.2]{dodis2004fuzzy}
 Let $A$, $B$, $C$ be random variables:
\begin{enumerate}[label=(\alph*)]
     \item For any $\delta > 0$, the conditional entropy $H_\infty(A|B = b)$ is at least $\tilde{H}_\infty(A|B)- log(1/\delta)$ with probability at least $1 -\delta$ over the choice of $b$.
     \item If $B$ has at most $2^n$ possible values, then $\tilde{H}_\infty(A | (B, C)) \geq \tilde{H}_\infty(A, B) | C)-\nu\geq \tilde{H}_\infty(A| C)-n$.
 \end{enumerate}
 \end{lemma}

A \textit{randomness extractor} maps a random variable with 
 guaranteed min-entropy, to a  uniformly distributed random variable from a smaller set such that
 the two variables have  small statistical distance. 
A {\em random source} is a random variable with  lower bound on its min-entropy.
We say  a  random variable $X$ defined over $\lbrace 0,1\rbrace^{n}$ is an {\em $(n,d)$-source}  if $H_\infty(X)\geq d$.
\begin{definition}\label{def: EXT}
A function $\ms{Ext}:\{0,1\}^n \times \mc S \to \lbrace 0,1 \rbrace ^{\ell}$ is a strong (seeded) $(d, \alpha)$ extractor if for any $(n,d)$-source $X$, we have: 
\[
\sd((S,\ms{Ext}(X,S));(S,U_{\ell})) \leq \alpha,
\]
where $S$ is chosen uniformly from $\mc S$. 
\end{definition}
  See \cite{nisan1996randomness} and references therein for more details. 
  A well known construction  of randomness extractors uses Universal Hash Families (UHF) whose randomness property is given by the  Leftover Hash
Lemma (LHL) \cite{impagliazzo1989pseudo}. We will use a variation of the LHL, called the \textit{generalized} LHL \cite[Lemma 2.4]{Dodis2008}. 
\begin{definition}[\text{Strong Universal Hash Family \cite{wegman1981new}}]\label{def: UHF}
A family of functions $\{\ms{h}_s:\mathcal{X}\to \mc Y\}_{s\in\mc S}$ is a  Strong Universal Hash Family if for any $x\neq x^{'}$ and any $a,b \in \mathcal{Y}$,
$
\mathrm{Pr}[h_S(x)=a \wedge h_S(x^{'})=b]= \frac{1}{|\mathcal{Y}|^2},
$
where the probability is over  the uniform choices over $\mathcal{S}$. 
\end{definition}

{
 \begin{lemma}[Generalized LHL]\label{lem:glhl}
 For two possibly dependant random variables $A\in \mathcal{X}$ and $B\in \mathcal{Y}$, 
 applying a universal hash function (UHF) $\{\ms{h}_s:\mathcal{X}\to \{0,1\}^\ell\}_{s\in\mc S}$ on $A$ 
 can extract  a uniformly random variable whose length $\ell$ will be bounded by the average min-entropy of $A$, given $B$, and the required closeness to the  uniform
 distribution. That is: 
\begin{align*}
  \sd(B,S,(h_{S}({A}));(B,S,U_\ell))\leq \frac{1}{2}\sqrt{2^{\ell-\tilde{H}_\infty(A|B)}},
\end{align*}
where $S$ is the randomly chosen seed of the hash function family, and
the average conditional min-entropy is defined above.
 \end{lemma}

}

\subsection{One-way Secret Key Agreement (OW-SKA)} 
One-way secret key agreement  in  source model   
was first considered  by Ahlswede \cite{Ahlswede1993}.  
Ahlswede considered  source model   
 where 
 Alice and Bob 
 have 
 instances  of the  initial correlated RVs 
  $X$ and $Y$, 
  and Eve has the same number of instances from their side-information $Z$, 
  and variables are obtained through a
joint public distribution $P_{XYZ}$, and defined ``Forward key capacity'' of key establishment protocols in which Alice transmits a single message to Bob. Such peotocols are later 
   called ``one-way secret key agreement''  (OW-SKA) \cite{Holenstein2006}. 
 


A one-way secret-key agreement protocol is defined using three parameters: a security parameter
 $\lambda$,  the secret key length $\ell$, and the number of instances of the initial RVs used $n$. Following \cite{Holenstein2006}, we assume for a given $\ell$ and $\lambda$, $n$ can be computed by a function $\ms n(\lambda,\ell)$.

\begin{definition}[
OW-SKA Protocol \cite{Holenstein2006}]\label{def:owska} Let $X$ and $Y$ be RVs defined over $\mc X$ and $\mc Y$, respectively.
For  the security parameter $\lambda$ and the  shared key length $\ell$  ($\lambda,\ell\in\mathbb{N}$),
%
a one-way secret-key agreement (OW-SKA)
protocol consists of a  function $\ms n:\mathbb{N}\times\mathbb{N}\to \mathbb{N}$ that specifies $n=\ms n(\lambda,\ell)$;
 a 
(probabilistic) function family  $\{{\tau}_{Alice}: \mc X^n\to \mc \{0,1\}^\ell\times \mc C \}_{\lambda,\ell}$,  mapping  $n$ instances of $X$ 
to a bit string $k_A\in\{0,1\}^\ell$ (the  secret key) and $C$  (the communication); and a function family $\{{\tau}_{Bob}: \mc Y^n\times \mc C\to \{0,1\}^\ell \}_{\lambda,\ell}$, mapping $c\in \mc C$ and $n$ instances of $Y$ to a bit
string $k_B\in \{0,1\}^\ell$.

\end{definition}

The goal of secret-key agreement is to  establish 
a  
 key $k=k_A=k_B$ 
 that appears  uniformly random 
 to Eve. 
 
\begin{definition}[Secure OW-SKA Protocol]\label{def:owskasec}
A OW-SKA protocol on $\mc X \times \mc Y$ is
secure on a probability distribution family $\mc P_{XYZ}$ defined over $\mc X\times \mc Y\times \mc Z$  if for
$\lambda,\ell\in\mathbb{N}$, the OW-SKA protocol outputs a $(\epsilon(\lambda),\sigma(\lambda))$-Secret Key (in short $(\epsilon(\lambda),\sigma(\lambda))$-SK) $K$, an RV over  $\mc{K}$  that  
            satisfies the following reliability and security properties: 
            \begin{align}
                \text{(reliability)}\quad& \pr[{K_A=K_B=K}]\geq 1-\epsilon(\lambda), \\
                \text{(security)}\quad& \sd\left((K,C ,Z);(U_{\{0,1\}^{\ell(\lambda)}},C,Z^n)\right) \leq \sigma(\lambda),\label{def:sec}
	        \end{align}
	        
 where $K_A$ and $K_B$ are random variables corresponding to $\tau_{Alice}(\cdot)$ and $\tau_{Bob}(\cdot)$ functions respectively, $n=\ms n(\lambda,\ell)$ and $\eps(\lambda)$ and $\sigma(\lambda)$ are small non-negative values. 
\end{definition}

\subsection{Hybrid Encryption}
 A  {\em hybrid encryption} scheme  is a public-key encryption (PKE) schcme that uses (i) a special PKE, known as KEM, that is used to encrypt a symmetric key that is decrytable by Bob and
 establishes a shared key between Alice and Bob, and  (ii)
  a symmetric key encryption scheme called DEM, that encrypts an arbitrarily long message.  
  
In the following, we use  $\lambda$ to denote a parameter  that determines the security parameter of the system, and use the unary representation $1^\lambda$  that is commonly used in cryptography\footnote{Unary representation of  
$\lambda \in  \mathbb{N}$, is a  bit string consisting of $\lambda$  copies of the 1.}.
 In the rest of this section the attacker is assumed to be computationally bounded.

\remove{Security of public and symmetric key parts of hybrid encryption systems must be defined such that the required security for the hybrid encryption system is achieved.
REPEAT
 Cramer et al  \cite{cramer2003design} and  Shoup \cite{shoup2001proposal} formalised  
{\em KEM/DEM paradigm} for hybrid encryption schemes,
defined adaptive CCA security for KEM and DEM,  and proved that if both KEM and  DEM are CCA2 secure, the resulting hybrid encryption will be CCA2  secure. 
Necessary and sufficient conditions on the security of the KEM and the DEM in order to guarantee a hybrid PKE scheme with a certain given level of security {\bl security parameter?)} is studied in \cite{herranz2006kem}.
}

\begin{definition}[Key Encapsulation Mechanism (KEM) \cite{shoup2001proposal}]\label{def:kem} 
A KEM $\kem =  (\gkem,\ekem, \dkem)$ with a security parameter $\lambda$ and a key space 
$\{0,1\}^{\lkem(\lambda)}$,  is a triple of algorithms 
 defined as follows:
 
\begin{enumerate}
    \item $\gkem(1^{\lambda})$ is a
  randomized  key generation algorithm that  takes the security parameter  $\lambda \in \mathbb{N}$ returns a public and secret-key pair $(pk, sk)$.
    \item  
    $\ekem(1^\lambda,pk)$ takes a public-key $pk$ and outputs a 
ciphertext $c$, and 
a key  $k\in\{0,1\}^{\lkem(\lambda)}$.
\item 
 $\dkem(1^\lambda,sk, c)$  is a  deterministic decapsulation algorithm that takes 
 a secret key $sk$ and a ciphertext $c$, and returns a key $k \in\{0,1\}^{\lkem(\lambda)}$, or $\perp$ that denotes failure,
\end{enumerate}
where private and public-key spaces are $\mathcal{SK}$ and $\mathcal{PK}$, respectively, and a ciphertext space is $\mathcal{C}$. That is $sk\in\mc{SK}$, $pk\in\mc{PK}$, and $c\in \mc C$.
\end{definition}

 A KEM $\kem$ is {\em $\eps(\lambda)$-correct} if for  all $(sk, pk) \leftarrow\gkem(1^{\lambda})$ and $(c, k)\leftarrow\ekem(1^\lambda,pk)$, it holds that $\text{Pr}[\dkem(1^\lambda,sk, c)\neq k]\leq \eps(\lambda) $,  where probability is over the choices of $(sk, pk)$ and  the randomness of $\ekem(\cdot)$ and $\eps(\cdot)$ is a non-negative negligible function in $\lambda$.

\textit{Security of KEM} is defined as the indistinguishability of the established key from a random string for an 
 attacker that may have access   to 
 decryption algorithm (attacker can always access the encryption algorithm using the public-key). 
A ciphertext query to the   {\em  decryption oracle}  
 (decryption  algorithm with correct private decryption key) 
results in 
the corresponding plaintext, or $\perp$ that denotes the ciphertext is not valid. This is  known as Chosen Ciphertext Attack (CCA) 
attack. 
In CCA1  security ciphertext queries are not allowed after the challenge ciphertext is seen, while in CCA2 security  such queries are allowed.
\textit{Chosen Plaintext Attack (CPA)} in KEM limits the attacker to encryption queries. 
 Since the attacker can freely encrypt any message because of their access to the public-key, CPA security for KEM implies the attacker has no query access.
We use notations and formalization of  \cite{herranz2006kem}. For compact representation of various attacks, let $ATK$  be a formal symbol that can take values $CPA;CCA1;CCA2$.
The \underline{K}ey \underline{ind}istinguishability  ($kind$) of a KEM  $\kem$  is formalized by bounding 
advantage  
$Adv^{kind}_{\kem,\msa}$ of an adversary $\msa$  as defined below.

\begin{definition}[Security of KEM: IND-CPA, IND-CCA1, IND-CCA2 \cite{herranz2006kem}] Let $\kem =  (\gkem,\ekem, \dkem)$ be a KEM scheme
and let $\msa = (\msa_1, \msa_2)$ be an adversary. For $atk \in\{cpa, cca1, cca2\}$ and $\lambda\in \mathbb{N}$,
\small{
\begin{align*}
    &Adv^{kind\text{-}atk}_{\kem,\msa}(\lambda) \triangleq |\pr[(pk,sk)\stackrel{\$}\gets\gkem(1^\lambda); st\stackrel{\$}\gets {\msa_1}^{\ms O_{1}}(1^\lambda,pk);\\
   &\qquad\qquad\qquad\qquad\qquad(k^*,c^*)\stackrel{\$}\gets \ekem(1^\lambda,pk);
    k_0\gets k^*;\\ &\qquad\qquad\qquad\qquad\qquad k_1\stackrel{\$}\gets\{0,1\}^{\lkem(\lambda)}; b\stackrel{\$}\gets \{0,1\};\\
    &\qquad\qquad\qquad\qquad\qquad\msa_2^{\ms O_{2}}(1^\lambda,c^*,st,k_b)=b]-\frac{1}{2}|, 
\end{align*}}
\normalsize

where

\small{
\begin{center}
\begin{tabular}{l l l}
    $atk$ & \ $\ms O_{1}(\cdot)$ & $\ms O_{2}(\cdot)$ \\
    \hline
     $cpa$& \ $\ \varepsilon$ & $\varepsilon$\\
     $cca1$& \ \ $\dkem_{sk}(\cdot)$ & $\varepsilon$\\
     $cca2$& \ \ $\dkem_{sk}(\cdot)$ \ \ &$\dkem_{sk}(\cdot)$\\
\end{tabular}
\end{center}
}

\normalsize
\vspace{1.2em}
Let $ATK\in\{CPA,CCA1,CCA2\}$. A {\em KEM is $\sigma(\lambda)\text{-}IND\text{-}ATK$ secure, } if for \textit{all} computationally bounded adversaries $\msa$, $Adv^{kind\text{-}atk}_{\kem,\msa}(\lambda)\leq \sigma(\lambda)$, where $\sigma(\cdot)$ is a non-negative negligible function of $\lambda$.
\end{definition}
{In above, $(c^*,k^*)$ are 
the \textit{challenge ciphertext and key} pair. We use $\ms{O}_{1}(\cdot)$ and $\ms{O}_{2}(\cdot)$ to denote (non-free) decapsulation oracles that can be accessed by the adversary before and after seeing the challenge output, respectively,    
$\dkem_{sk}(\cdot)$ is the decapsulation oracle with 
 private key $sk$, and $\varepsilon$ denotes an empty string.}

\begin{definition}
[Data Encapsulation Mechanism (DEM) \cite{shoup2001proposal}]\label{def:dem} A DEM $\dem=(\gdem,\edem,\ddem)$ with  security parameter $\lambda$ and 
a  key space $\{0,1\}^{\ldem(\lambda)}$  consists of two algorithms:
\begin{enumerate}
\item $\gdem(1^\lambda)$ is the randomized key-generation algorithm produces a uniformly distributed key $k\in\{0, 1\}^{\ldem(\lambda)}$. 
  \item $\edem(1^\lambda,k,m)$ is the randomized encryption algorithm that encrypts message $m$ under the uniformly chosen key $k\in\{0,1\}^{\ldem(\lambda)}$ and outputs a ciphertext $c$.
  \item $\edem(1^\lambda,c,k)$ is the deterministic  decryption algorithm that decrypts the ciphertext $c$ using the key $k$ to get back a message $m$ or the special rejection symbol $\perp$.
\end{enumerate}
\end{definition}

\remove{
Similar to \cite{shoup2001proposal}, we  assume 
encryption and decryption algorithms are deterministic, and that the
scheme is (perfectly) correct (i.e. ``sound'' in the terminology of \cite{shoup2001proposal}),  and 
  {for all $k\in\{0,1\}^{\ldem(\lambda)}$, 
  $m \in \{0,1\}^* $, we have $\pr [\ddem\big(1^\lambda,k,\edem(k,m)\big) = m] = 1$.}}

\textit{Security of DEM} 
 against CPA, CCA1, and CCA2  is defined in
  \cite{shoup2001proposal} and  is 
   the same as the corresponding definitions 
   for symmetric encryption schemes as defined in \cite{bellare1997concrete}. 
 DEM is a symmetric key primitive and so unlike KEM 
 access to encryption oracle is a resource.
{CPA security of DEM allows the attacker to have access to encryption oracle.}
Herranz et al. \cite{herranz2006kem} considered two one-time attacks for DEMs, known as one-time (OT) attack that is an attack without any access to the encryption oracle, and one-time chosen-ciphertext attack (OTCCA), where the attacker has access to chosen-ciphertext queries after observing the challenge,   and correspond to \textit{passive} and \textit{adaptive chosen ciphertext} attacks, respectively that were considered in \cite[Section 7.2.1]{cramer2003design}.
 These two security definitions are tailored to the application of hybrid encryption scheme and allow constructing the DEM part of the hybrid encryption from 
 a \textit{one-time symmetric encryption scheme} that can be realized by a block cipher for generating a pseudorandom sequence to be XORed with  the message. This scheme yields a DEM with OT security. By attaching a \textit{message authentication code} (MAC) to the one-time symmetric encryption scheme security against adaptive ciphertext attack is guaranteed \cite[Theorem 7.1]{cramer2003design}.
Security of a DEM $\dem$ is formalized by bounding the \underline{ind}istinguishability advantage $Adv^{ind}_{\dem,\msa}$ of an  adversary $\msa$, 
 defined as follows. 
%
\begin{definition}
[Security of DEM: IND-OT, IND-OTCCA, IND-CPA, IND-CCA1, IND-CCA2 \cite{herranz2006kem}]\label{def:demsec} Let $\dem =  (\edem, \ddem)$ be a DEM scheme with security parameter $\lambda$ and key space $\{0,1\}^{\ldem(\lambda)}$
and let $\msa = (\msa_1, \msa_2)$ be an adversary. For $atk \in\{ot,otcca, cpa, cca1, cca2\}$ and $\lambda\in \mathbb{N}$,  
\begin{align*}
    Adv^{ind\text{-}atk}_{\dem,\msa}(\lambda) \triangleq|&\pr[k\stackrel{\$}\gets\edem(1^\lambda);\\ &(st,m_0,m_1)\stackrel{\$}\gets \msa_1^{\ms O_1}(1^\lambda);
    b\stackrel{\$}\gets \{0,1\};\\
    &c^*\gets \stackrel{\$}\edem(1^\lambda,k,m_b);\msa_2^{\ms O_2}(1^\lambda,c^*,st)=b]\\
    &-\frac{1}{2}|,
\end{align*}
\normalsize
where

\footnotesize{
\begin{center}
\begin{tabular}{p{3em} p{11em} l}
    $atk$ &$\ms O_{1}$ &$\ms O_{2}$ \\
    \hline
     $ot$& $\varepsilon$ & $\varepsilon$\\
     $otcca$& $\varepsilon$&$ \ddem_k(\cdot)$\\
     $cpa$&$\edem_k(\cdot)$ &$\varepsilon$\\
     $cca1$&$\{\edem_k(\cdot),\ddem_k(\cdot)\}$& $\varepsilon$\\
     $cca2$&$\{\edem_k(\cdot),\ddem_k(\cdot)\}$& $\{\edem_k(\cdot),\ddem_k(\cdot)\}$\\
\end{tabular}
\end{center}
}

\normalsize
A DEM is $\sigma(\lambda)\text{-}IND\text{-}ATK$ for $ATK\in\{OT,OTCCA,CPA,CCA1,CCA2\}$ if for \textit{all} adversaries $\msa$, $Adv^{ind\text{-}atk}_{\dem,\msa}(\lambda)\leq \sigma(\lambda)$, where $\sigma(\cdot)$ is a non-negative negligible function in $\lambda$.
\end{definition}
{
Here, $\edem_k(\cdot)$ and $\ddem_k(\cdot)$ are  encryption and  decryption oracles  with 
 key $k$, respectively, and $\varepsilon$  denotes ``empty"; that is no oracle.} 
 {
\begin{definition}
[Hybrid PKE (HPKE) \cite{shoup2001proposal}]\label{def:hpke} An HPKE  $\hpk_{\kem,\dem}=(\ghpk,\ehpk,\dhpk)$ 
{is a public-key encryption algorithm} that uses  a pair of KEM $\kem=(\gkem,\ekem,\dkem)$ and DEM $\dem=(\edem,\ddem)$ algorithms with a  common  
key space $\{0,1\}^{\ell(\lambda)}$, and  consists of three algorithms for key generation, encryption and decryption defined in Fig.~\ref{fig:hpke}. 

\begin{figure}[h!]
   \begin{center}
\begin{tabular}{l| l }
  $\mathbf{Alg}\ \ghpk(1^\lambda)$ &  $\mathbf{Alg}\ \ehpk(1^\lambda,pk,m)$\\
      \small$(pk,sk)\stackrel{\$}\gets\gkem(1^\lambda)$&\small$(c_1,k)\stackrel{\$}\gets\ikeme(1^\lambda,pk)$\\
      \small Return $(pk,sk)$ &\small $c_2\gets\edem(1^\lambda,k,m)$\\
     &\small Return $(c_1,c_2)$
\end{tabular}
\begin{tabular}{l}
\\
$\mathbf{Alg}\ \ehpk(1^\lambda,sk,c_1,c_2)$\\
\small$k\gets\dkem(1^\lambda,y,c_1)$\\
\small If $\perp \gets\ikemd(1^\lambda,y,c_1)$: Return $\perp$\\
\small \quad Else:$m\gets\ddem(1^\lambda,c_2,k)$\\
\small \quad Return $m$
\end{tabular}
\end{center}
    \caption{Hybrid public-key encryption}
    \label{fig:hpke}
\end{figure}
\end{definition}
}

\remove{
\begin{remark}\label{remark:1}
A hybrid encryption scheme that consists of a (cryptographic) KEM and DEM schemes (e.g., the scheme of \cite[Section 7]{cramer2003design}) is a public-key encryption scheme and even though the access to the DEM's encryption algorithm individually is not free for an attacker, when the key of DEM is generated by  KEM, the adversary gets automatic access to the encryption algorithm through the public-key of the KEM and is able to encrypt messages of their choice by first generating a KEM key from the public-key and then using the key for data encryption in the DEM scheme.
\end{remark}
}

 The following composition theorem gives security of 
 hybrid encryption 
  \cite{herranz2006kem} (Theorem 5.1). 


\begin{theorem}[\textit{IND-ATK KEM + IND-ATK$'$ DEM $\Rightarrow$ IND-ATK PKE}]\cite[Theorem 5.1]{herranz2006kem}
] 
{
Let ATK$\in\{$CPA, CCA1, CCA2$\}$ and ATK$'\in\{$OT, OTCCA$\}$.
If ATK $\in\{$CPA, CCA1$\}$ and ATK$'$ = OT,  then the hybrid public-key encryption scheme $\hpk_{\kem,\dem}$ is a secure public-key encryption scheme under IND-ATK attack.
Similarly, if ATK = CCA2, ATK$'$ = OTCCA, $\hpk_{\kem,\dem}$ is a secure public-key encryption scheme under IND-ATK attack.
}

\remove{
For ATK$\in\{$CPA, CCA1, CCA2$\}$ and ATK$'\in\{$OT, OTCCA$\}$, if $\kem$ is a secure KEM under IND-ATK attacks and $\dem$ is a secure DEM under IND-ATK$'$ attacks, then the hybrid public-key encryption scheme $\hpk_{\kem,\dem}$ is a secure public-key encryption scheme under IND-ATK attack, where for ATK $\in\{$CPA, CCA1$\}$, ATK$'$ = OT and for ATK = CCA2, ATK$'$ = OTCCA.
}
\end{theorem}


{In Section~\ref{sec:new}, we prove a similar 
 composition theorem for iKEM and DEM for  specific security notions.}


\section{ Hybrid encryption in preprocessing model}\label{sec:new}
In the preprocessing model Alice, Bob and the attacker have access to their corresponding samples of a joint distribution that are generated during an \textit{offline phase}, and  prior to generating an encapsulated key. The distribution is public but the samples are {\em private inputs} of the parties. 
During the \textit{online phase} Alice and Bob use their private samples to {establish} a shared key and use as a symmetric key to encrypt their messages under a symmetric key encryption scheme.

A hybrid encryption in preprocessing model, denoted by $\ike_{\ikem,\dem}$, uses   a pair of algorithms $iKEM$ with information-theoretic security, and DEM with computational security, and results in a hybrid encryption with computational security.
We first define  \textit{information-theoretic KEM (iKEM)} and its security notions, and then describe the  $\ike_{\kem,\dem}$ system that uses a DEM  as defined in Definition~\ref{def:dem}.

\subsection{KEM in Preprocessing Model (iKEM)}\label{section:ikem}
 An iKEM allows Alice and Bob to use their samples of correlated randomness and a single message from Alice to Bob, to obtain a shared key that is secure against     an  eavesdropper (a wiretapper)
  with side information that is represented by their initial random samples.

\begin{definition}[iKEM]\label{def:ikem}
An iKEM $\ikem=(\ikemg,\ikeme,\ikemd)$ with security parameter $\lambda$, initial distribution $\mc P$, and the key space $\{0,1\}^{\likem_{\mc P}(\lambda)}$ is defined by a triple of algorithms  as follows:
 
\begin{enumerate}
    \item  $\ikemg(\mc P)$ the generation algorithm takes  a publicly known family of distributions $\mc{P}$ , and provides private inputs to Alice and Bob, and possibly Eve, denoted by $x,y$ and $z$, respectively.
    \item $\ikeme(x)$, the encapsulation algorithm, is a probabilistic algorithm that takes Alice's random string $x$ as input  and outputs a ciphertext $c$ and key $k\in\{0,1\}^{\likem_{\mc P}(\lambda)}$. 
    \item $\ikemd(y,c)$, the decapsulation algorithm, is a  deterministic algorithm that takes the receiver's random string $y$  and ciphertext $c$ as input, and outputs a  key $k$ or special symbol $\perp$ ($\perp$ that implies that the ciphertext was invalid).
\end{enumerate}
\end{definition}

\subsubsection*{Correctness of iKEM}  
{For  $\ikeme(x)= (c,k)$, let denote   $\ikeme(x).key= k$ and $\ikeme(x).ctxt=c$.}
The iKEM is $\eps(\lambda)$-correct if  for a 
 sample pair $(x,y)$,
$\text{Pr}[\ikemd(y, c)\neq \ikeme(x).{key}]\leq \eps(\lambda)$, where $\eps(\cdot)$
is a small function of $\lambda$ and
 probability is over  all the random coins of $\ikeme$,  $\ikemd$ and $\ikemg$.


 
\subsubsection*{Security of  iKEM}
Security of iKEM is against a computationally unbounded attacker  
that has the side information $z$, and can query the encapsulation and decapsulation algorithms.  
For  sampled private input pair $(x;y)$, we define iKEM encapsulation $\ikeme_x(\cdot)$ and decapsulation $\ms \ikemd_y(\cdot)$ oracles, and use them to define Chosen Encapsulation Attack (CEA)\footnote{We note that in chosen encapsulation attack, the attacker doesn't actually make any choice regarding the content of its attack (like in chosen plain/ciphertext attacks), rather, choose to reach the encapsulation oracle and query it.} and Chosen Ciphertext Attack
(CCA), respectively.

\remove{
We thus, consider two types of oracles for a sampled private input pair $(x,y)$, namely,iKEM enacapsulation $\ikeme_x(\cdot)$ and iKEM decapsulation $\ms \ikemd_y(\cdot)$, and their corresponding  attacks, 
{\em Encapsulation Oracle Attack (CEA)} and  {\em Chosen Ciphertext Attack (CCA)},.}

A query to  $\ikeme_x(\cdot)$  does not have any input, and  outputs a pair $(c,k)$ where $k$ and $c$ are a key and the corresponding ciphertext 
 that is obtained by using the secret input of Alice and other system's public information.
A query to $\ms \ikemd_y(\cdot)$, is a 
ciphertext  $c$ that is chosen by the attacker, and will result in  
the output either a key $k$, or $\perp$, indicating that $\ikemd$ can/cannot generate a valid key for the presented $c$.
 
 %
\remove{
 An established key will use the entropy of the correlated variables that is sampled before the protocol starts which will be reduced with each query.   
 We  thus {\rd consider security against a computationally unbounded adversary with access to a \textit{bounded} number of queries (because otherwise a computationally unbounded adversary can keep asking the oracles until all information about the encrypted message is leaked). In particular,}
}
We consider three types of attackers: an attacker with no access to encapsulation or decapsulation oracles (OT attack), 
 an attacker with  access to  $q_e$ encapsulation queries ($q_e\text{-}CEA$ attack), 
and an  attacker that has access to  {$q_e$ encapsulation} and $q_d$ decapsulation queries  ($(q_e;q_d)\text{-}CCA$ attack).  
 The corresponding {security notions  
 are  denoted by IND-OT,  
 IND-$q_e$-CEA, and IND-$(q_e;q_d)$-CCA,}  respectively. 
  For a given security parameter $\lambda$, and an input family of distributions $\mc P$, the number of queries affect the maximum key length that can be established using iKEM. In particular, more queries result in shorter keys. Note that this also enforces an upper-bound on the number of queries to allow achieving a positive key length.
We use   $\advu = (\advuone,  \advutwo)$
 to denote  an adversary with  ``U''nbounded  computation   that  uses algorithm $\advuone$  before seeing the challenge, and  passes the learnt information (its state) to algorithm $\advutwo$  that is executed 
 after seeing the challenge. 
{Security of an  iKEM $\ikem$ is formalized by bounding the \underline{i}nformation-theoretic \underline{k}ey \underline{ind}istinguishability (ikind) advantage  $Adv^{ikind}_{\ikem,\advu}$  of an  adversary $\advu$, and is defined as follows. }
\begin{definition} 
[Security of iKEM: IND-OT, 
IND-$q_e$-CEA , IND-$(q_e;q_d)$-CCA] \label{def:sikem} Let $\ikem =  (\ikemg,\ikeme, \ikemd)$ be an iKEM scheme with security parameter $\lambda$, input family of distributions $\mc P$, and the key space $\{0,1\}^{\likem_{\mc P,q}(\lambda)}$ 
and let $\advu = (\advuone, \advutwo)$ be an unbounded adversary. For $\lambda\in \mathbb{N}$, $atk\in\{ot,  q_e\text{-}cea,  
(q_e;q_d)\text{-}cca\}$, and  $q\in\{null,q_e,(q_e;q_d)\}$, respectively,  define, 
%
\small
\begin{align*}
    Adv^{ikind\text{-}atk}_{\ms {iK},\advu}(\lambda) \triangleq|\pr[&(x,y,z)\stackrel{\$}\gets\ikemg(1^\lambda,\mc P); st\stackrel{\$}\gets \sadvuone^{\ms O_{1}}(z);\\
   &(k^*,c^*)\stackrel{\$}\gets \ikeme(1^\lambda,x);
    k_0\gets k^*;\\ &k_1\stackrel{\$}\gets\{0,1\}^{\likem_{\mc P,q}(\lambda)};b\stackrel{\$}\gets \{0,1\};\\ &\sadvutwo^{\ms O_{2}}(st,c^*,k_b)=b]-\frac{1}{2}|,
\end{align*}
\normalsize
where 
\footnotesize
\begin{center}
\begin{tabular}{l l l}
    $atk$& $\ms O_{1}(\cdot)$ & $\ms O_{2}(\cdot)$\\
    \hline
     $ot$& $\varepsilon$ & $\varepsilon$\\
         $q_e\text{-}cea$  
          &  $\ikeme_x(\cdot)$ & $\varepsilon$\\
     $(q_e;q_d)\text{-}cca$&  $\{\ikeme_x(\cdot),\ikemd_y(\cdot)\}$  &$\{\ikeme_x(\cdot),\ikemd_y(\cdot)\}$\\
\end{tabular}
\end{center}
\normalsize

An iKEM is $\sigma(\lambda)\text{-}IND\text{-}ATK$   secure  for  $ATK\in\{\text{OT}, 
q_e\text{-CEA}, (q_e;q_d)\text{-CCA}\}$,  if  for \textit{all} adversaries $\advu$, $Adv^{ikind\text{-}atk}_{\ikem,\advu}(\lambda)\leq \sigma(\lambda)$, where $\sigma(\cdot)$ is a non-negative small function of $\lambda$.
\end{definition}
{In above, $\ms O_{1}$ and $\ms O_{2}$ are oracles that can be accessed before and after receiving the challenge ciphertext, respectively.}

The following lemma shows that the distinguishing advantage of the adversary $\advu$ in Definition~\ref{def:sikem} is bounded by the the statistical distance  of the generated key with uniform distribution, given adversary's view of the game. {This lemma can be seen as a special case of \cite[Lemma 4]{maurer2007indistinguishability}, where the \textit{random system} is an iKEM.} 


 Let $\mathbf{v}^{q_e\text{-}cea} =( {v_1}^{cea}, \cdots, {v_{q_e}}^{cea})$, where for $1\leq i\leq q_e$, $v^{cea}_i\in \{0,1\}^{\likem_{\mc P,q_e}(\lambda)}\times \mc C$, denote the encapsulation oracle responses to the adversary's queries in a  $q_e$-bounded CEA attack,
and $\mathbf{V}^{q_e\text{-}cea}$ denote the corresponding random variable
{(i.e. probabilistic view due to the  
 $\ikeme$ random coins, and  private sample of Alice).} We note that since the encapsulation oracle does not take any input from the adversary, $v^{q_e\text{-cea}}$ does not depend on the adversary and is the same for all adversaries. 
\begin{lemma}\label{lem:sdeno}
An  iKEM $\ikem$ is $\sigma(\lambda)$-indistinguishable against  $q_e$-bounded CEA, 
{ \em if and only if} for  all adversaries $\advu$, we  have
\small
\begin{align}  \label{eq:ikemsd}
\sd\big(&(Z,C^*,K^*,\mathbf{V}^{q_e\text{-}cea});(Z,C^*,U_{\likem_{\mc P,q_e}(\lambda)},\mathbf{V}^{q_e\text{-}cea})\big)\leq \sigma(\lambda),
\end{align}
\normalsize
{where  random variables $Z$,  and $(C^*, K^*)$ correspond to $z$, the attacker's initial side information, 
 and the pair $(c^*,k^*)$ of the challenge ciphertext and key pair,  respectively.}

\end{lemma}

\textit{Proof of Lemma~\ref{lem:sdeno}}. The proof has two  directions: 
 (a) the iKEM  is indistinguishable if the statistical distance is bounded, and (b) if the iKEM is indistinguishable then the statistical distance is bounded.
 

(a) Consider 
an iKEM that is $\sigma(\lambda)$-IND-$q_e\text{-CEA}$ secure according to Definition~\ref{def:sikem}. 
If  (\ref{eq:ikemsd})  does not hold, 
 there exists  a set $\mc W \subset \mc Z\times ( \{0,1\}^{\likem_{\mc P,q_e}(\lambda)}\times \mc C)^{q_e+1}$ (note that $q_e+1$ corresponds to $q_e$ pair of queried keys and ciphertexts and \textit{one} challenge pair) for which 
\begin{align*}
    |\pr[&\big((Z,K^*,C^*,\mathbf{V}^{q_e\text{-}cea})\in\mc W\big)]\\
    &-\pr[\big(Z,U_{\likem_{\mc P,q_e}(\lambda)},C^*,\mathbf{V}^{q_e\text{-}cea})\in\mc W\big)])>\sigma(\lambda)
\end{align*}

We use $\mc W$ to define an adversary algorithm  $\advbu=(\advbuone,  \advbutwo)$ for iKEM security experiment (in Definition~\ref{def:sikem})  that for all side information $z$, challenge pair $(c^*,k^*)$, and $q_e$ encapsulation oracle outputs $\mathbf{v}^{q_e\text{-}cea}$ that satisfies  $(z, c^*,k^*,\mathbf{v}^{q_e\text{-}cea})\in \mc W$,  outputs  zero (that is  chooses $k_0$ as its response).  
 This allows $\advbu$ to gain an advantage $Adv^{ikind\text{-}atk}_{\ms \ikem,\advbu}(\lambda)>\sigma(\lambda)$, and this contradicts the assumption (that the iKEM is $\sigma(\lambda)$-indistinguishable). Therefore the statistical distance is less than $\sigma(\lambda)$.

(b) Suppose (\ref{eq:ikemsd}) holds, then we define an adversary $\advbu=(\advbuone,\advbutwo)$ that 
its output at the end of the probabilistic experiment (in Definition~\ref{def:ikem}) defines a function $\ms {F_{\advbu}}:\mc Z\times(\{0,1\}^{\likem_{\mc P,q_e}(\lambda)}\times \mc C)^{q_e+1} \to \{0,1\}$
that takes $\advbu$'s input $z,c^*$, $k^*$ and $\mathbf{V}^{q_e\text{-}cea}$, and outputs 0 or 1.  Then we have:
\begin{align*}
    &Adv^{ikind\text{-}atk}_{\ikem,\advu}(\lambda)\leq \\
    &\qquad\displaystyle\max_{\ms {F_{\advu}}}|\pr[\ms {F_{\advu}}(Z,C^*,K^*,\mathbf{V}^{q_e\text{-}cea})= 1]\\
    &\qquad\qquad-\pr[\ms {F_{\advu}}(Z,C^*,U_{\likem_{\mc P,q_e}(\lambda)},\mathbf{V}^{q_e\text{-}cea})= 1].
\end{align*}
Let $\mc W \subset \mc Z\times(\{0,1\}^{\likem_{\mc P,q_e}(\lambda)}\times \mc C)^{q_e+1} $ be the set for which  $(\pr[\big((Z,C^*,K^*,\mathbf{V}^{q_e\text{-}cea})\in \mc W\big)]-\pr[\big((Z,C^*,U_{\likem_{\mc P,q_e}(\lambda)},\mathbf{V}^{q_e\text{-}cea})\in \mc W\big))])$ is maximized.
Then consider an adversary that outputs 1 only when  $(z,c^*,k^*,v^{q_e\text{-}cea})\in \mc W$. This corresponds to ${\ms {F_{\advbu}}}$ to be non-zero when $(z,c^*,k^*,v^{q_e\text{-}cea})\in\mc W$. From the definition of the \textit{statistical distance} (\ref{eq:sd}), it implied that 
\begin{align*}
    &Adv^{ikind\text{-}atk}_{\ikem,\advu}(\lambda)\leq \displaystyle\max_{\ms {F_{\advbu}}}|\pr[\ms {F_{\advbu}}(Z,C^*,K^*,\mathbf{V}^{q_e\text{-}cea})= 1]\\
    &\qquad\qquad\qquad\qquad-\pr[\ms {F_{\advbu}}(Z,C^*,U_{\likem_{\mc P,q_e}(\lambda)},\mathbf{V}^{q_e\text{-}cea})= 1]\\
    &=\sd\big((Z,C^*,K^*,\mathbf{V}^{q_e\text{-}cea});(Z,C^*,U_{\likem_{\mc P,q_e}(\lambda)},\mathbf{V}^{q_e\text{-}cea})\big)\\
    &\leq \sigma(\lambda)\ \ \blacksquare
\end{align*}

\begin{corol} \label{cor}The iKEM in Definition~\ref{def:ikem} is IND-OT secure if and only if:
\begin{equation}\label{eq:cpaikemsd}
\sd\big((Z,C^*,K^*);(Z,C^*,U_{\likem_{\mc P,q_e}(\lambda)})\big)\leq \sigma(\lambda),
\end{equation}
{where  random variables $Z$,  and $(C^*,  K^*)$ correspond to $z$, and the pair of  challenge ciphertext and key 
 $(c^*,k^*)$, }respectively.
\end{corol}
\textit{Proof.} The proof follows from Lemma~\ref{lem:sdeno} and noting that for IND-OT security {no}  query is allowed for the adversary and $\mathbf{v}^{q_e\text{-}cea}_{\ms A}$ is empty. $\ \blacksquare$

\subsection{DEM in Preprocessing Model}

{Hybrid encryption in preprocessing model will use 
 {Definition~\ref{def:dem}  for DEM with security notions as defined  in Defintion~\ref{def:demsec} 
  against 
 a computationally ``B"ounded adversary} that will be denoted by  $\msa^{\ms B}$.

}

\subsection{Hybrid Encryption using iKEM}

{Hybrid encryption in preprocessing model  uses private  samples of correlated variables as the key material
in an iKEM  with information-theoretic security against (an unbounded attacker) $\msa^{\ms U}$, and a DEM with computational security against  (a bounded) attacker $\msa^{\ms B}$), and provides a computationally secure encryption system. }

\begin{definition}\label{def:he}
[Hybrid Encryption (HE) in Preprocessing Model] For a security parameter $\lambda\in \mathbb{N}$ and an input family of distributions $\mc P$, 
let $\ikem=(\ikemg,\ikeme;\ikemd)$ and $\dem=(\edem,\ddem)$ be a pair of iKEM and  DEM  defined for the same security parameter, and
{the same key space for each $\lambda$.}
We define a
{\em hybrid encryption in preprocessing model} denoted by $\ike_{\ikem,\dem}= (\gike,\eike,\dike)$ using
an iKEM and a DEM, as in Fig.~\ref{fig:gkemcomb}.
\begin{figure}[h!]
   \begin{center}
\begin{tabular}{l| l }
  $\mathbf{Alg}\ \gike(\mc P)$ &  $\mathbf{Alg}\ \eike(x,m)$\\
      \small$(x,y,z)\stackrel{\$}\gets\ikemg(\mc P)$&\small$(c_1,k)\stackrel{\$}\gets\ikeme(x)$\\
      \small Return $(x,y,z)$ &\small $c_2\gets\edem(k,m)$\\
     &\small Return $(c_1,c_2)$
\end{tabular}
\begin{tabular}{l}
\\
$\mathbf{Alg}\ \dike(y,c_1,c_2)$\\
\small$k\gets\ikemd(y,c_1)$\\
\small If $\perp \gets\ikemd(y,c_1)$:Return $\perp$\\ 
\small \quad Else: $m\gets\ddem(c_2,k)$\\
\small \quad Return $m$
\end{tabular}
\end{center}
    \caption{Information-theoretic hybrid encryption}
    \label{fig:gkemcomb}
\end{figure}
\end{definition}
A hybrid encryption scheme in preprocessing
model uses private samples and  so  access to the encryption
oracle (CPA) means that  the attacker sees the output of the encryption system on messages of its choice using the same private sample.

Depending on the attacker’s access to the encryption and decryption oracles, we consider three security notions.  In  
computational setting,  the maximum number of queries  is a polynomial function of the security parameter. In information-theoretic security, the maximum number of  queries is a function of the security parameter  and the input family of distributions, such that achieving a positive key length is guaranteed. For a given key length,  we  thus consider security of iKEM   for a fixed number of queries.
Fixed number of queries (for given security parameter) has also been considered in computational setting \cite{cramer2007bounded,fuller2015unified}
to overcome some impossibility results that hold for general encryption schemes\footnote{For example, indistinguishability  CCA security from indistinguishability CPA security cannot be achieved without extra assumption.}. 
We define one-time CPA attack, denoted by IND-OT, for an attacker with no oracle access (passive attacker), inline with OT
attack in DEM.  We also define IND-$q_e$-CPA where the attacker has access to a fixed number of encryption queries
to to $\eike_x(\cdot)$ oracle.  Decryption queries will be defined similar to that of HPKE for the decryption
oracle $\dike_y(\cdot)$.
We define IND-$(q_e; q_d)$-CCA security of a hybrid encryption in preprocessing model where the attacker has access to $q_e$ encryption and $q_d$ decryption queries and
the oracles can be $\eike_x(\cdot)$, or $\dike_y(\cdot)$ (encryption and
decryption oracles) as follows.

\begin{definition}\label{def:she}
[IND-OT, IND-$q_e$-CPA, IND-$(q_e;q_d)$-CCA security of hybrid 
encryption in preprocessing model] Let $\ike_{\ikem,\dem}= (\gike,\eike,\dike)$ be a hybrid encryption in preprocessing model using 
 an iKEM $\ikem=(\ikemg,\ikeme;\ikemd)$ 
and a DEM $\dem=(\edem,\ddem)$, and 
Let $\advb = (\advbone, \advbtwo)$ be a computationally bounded adversary. For  $atk \in\{ ot, q_e\text{-}cpa, (q_e;q_d)\text{-}cca\}$ and $\lambda\in\mathbb{N}$, define 
\begin{align*}
   Adv^{ind\text{-}atk}_{\ike,\advb}(\lambda) \triangleq &|\pr[(x,y,z)\gets\gike(\mc P);\\
   &(st,m_0,m_1)\stackrel{\$}\gets \sadvbone^{\ms O_{1}}(1^\lambda,z);b\stackrel{\$}\gets \{0,1\};\\
   &c^*\stackrel{\$}\gets  \eike(x,m_b);\sadvbtwo^{\ms O_{2}}(1^\lambda,st,c^*)=b]\\
    &-\frac{1}{2}|,
\end{align*}
where
\footnotesize
\begin{center}
\begin{tabular}{l l l}
    $atk$& \ $\ms O_{1}(\cdot)$ & $\ms O_{2}(\cdot)$\\
    \hline
     $ot$& \ $\ \varepsilon$ & $\varepsilon$\\
     $q_e\text{-}cpa$&$\eike_x(\cdot)$ & $\varepsilon$\\
     $(q_e;q_d)\text{-}cca$&$\{\eike_x(\cdot),\dike_y(\cdot)\}$&$\{\eike_x(\cdot),\dike_y(\cdot)\}$\\
\end{tabular}
\end{center}
\vspace{1.2em}
\normalsize
A hybrid encryption  scheme $\ike_{\ikem,\dem}$ in preprocessing model is $\sigma(\lambda)\text{-}IND\text{-}ATK$ {for} $ATK\in\{\text{CPA},q_e\text{-CPA},(q_e;q_d)\text{-CCA}\}$ if  for \textit{all} adversaries $\advb$, $Adv^{ind\text{-}atk}_{\ike,\advb}(\lambda)\leq \sigma(\lambda)$, where $\sigma(\cdot)$ is a non-negative small function of $\lambda$.
\end{definition}

{In above, $\ms O_{1}$ and $\ms O_{2}$ are (non-free) oracles that can be accessed before and after receiving the challenge ciphertext, respectively.}

The following
composition theorem for hybrid encryption
shows that, an $\text{IND-OT}$ secure iKEM and an IND-OT secure DEM gives
an $\text{IND-OT}$ secure HE, and a $q_e$-\text{CEA} secure iKEM and an IND-OT secure DEM gives
a $q_e$-\text{CPA} secure HE.

\begin{theorem}[\textit{IND-$q_e\text{-CEA}$ iKEM + IND-OT$'$ DEM $\Rightarrow$ IND-$q_e\text{-CPA}$ HE}]\label{theo:composition} 
Let $\ms{iK} $ denote an iKEM with security parameter $\lambda$ that is  $\sigma(\lambda)$-IND-$q_e\text{-CPA}$ secure (information theoretically secure), and 
$\dem$ denote a   $\sigma'(\lambda)$-IND-OT secure (computationally secure) DEM with a security parameter $\lambda\in \mathbb{N}$,  and  assume  $\ikem$ and $\dem$ have  compatible key spaces  
 $\{0,1\}^{\ell_{\mc P,q_e}(\lambda)}$.
Then, the hybrid encryption scheme $\ms{HE}_{\ikem,\dem}$ is  a computationally secure  IND-$q_e\text{-CPA}$ secure hybrid encryption in preprocessing model with  security  against a computationally bounded adversary $\advb=(\advbone,\advbtwo)$.
\remove{
for  ATK$''\in \{q_e\text{-CPA},(q_e;q_d)\text{-CCA}\}$,
{where 

(i)  for ATK$=$ and  ATK$'=OT$,  we have ATK$''=q_e\text{-CPA}$ and 
$$Adv^{ind\text{-}q_e\text{-}cpa}_{\ike,\advb}(\lambda)\leq \sigma(\lambda)+\sigma'(\lambda),$$

(ii)  for $ATK=(q_e;q_d)\text{-CCA}$ and  ATK$'=OTCCA$, we have ATK$''=(q_e;q_d)\text{-CCA}$ and 
$$Adv^{ind\text{-}(q_e;q_d)\text{-}cca}_{\ike,\advb}(\lambda)\leq \eps(\lambda)+\sigma(\lambda)+\sigma'(\lambda).$$}}
\end{theorem}

\textit{Proof.} 
We prove the claim of the theorem for the second case that is, IND-$q_e$-CEA secure iKEM and an IND-OTCCA secure DEM results in an IND-$q_e$-CPA secure hybrid encryption scheme. 
The proof of the former case will use a similar argument. We define two consecutive experiments that models the adversary’s
interaction with the encryption system and its modified version, respectively. Both experiments operate on the same underlying probability space. In particular, private inputs of parties, randomness of the adversary's  algorithm, and the hidden bit $b$ take on identical values across all experiments. At the end of each experiment, the adversary outputs a bit $\hat{b_i}$, where $i$ corresponds to the index of the experiment. 

The two experiments are as follows: \textit{Experiment 0} denoted by ``EXP-0'' is identical to the experiment used in the security definition of hybrid encryption in preprocessing model (Definition~\ref{def:she}) defined as
\begin{align*}
&\text{EXP-0}\triangleq\\
&[(x,y,z)\stackrel{\$}\gets\gike(\mc P);(st,m_0,m_1)\stackrel{\$}\gets \sadvbone^{\ms O_1}(1^\lambda,z);\\ & b\stackrel{\$}\gets \{0,1\}; c^*\stackrel{\$}\gets \eike(1^\lambda,k,m_b);{\hat{b}_0\gets\sadvbtwo^{\ms O_2}}(1^\lambda,c^*,st)],\end{align*}



 \textit{Experiment 1} denoted by ``EXP-1'' is defined with regards to EXP-0 and only differs from it in using a uniformly sampled key instead of the key generated by iKEM for encryption and answering encryption  queries (by oracles $\ms O'_1$ and $\ms O'_2$).
 
 For an experiment EXP-$i$, where $i\in\{0,1,\}$ with output $\hat{b_i}$, $\pr[T_i]$  denotes the event that  $\hat{b_i}=b$.




\remove{
\begin{align*}
&\text{EXP-2}\triangleq\\
&\qquad[(x,y,z)\stackrel{\$}\gets\gike(1^\lambda);k\stackrel{\$}\gets\{0,1\}^{\ell(\lambda)};\\
& \qquad(st,m_0,m_1)\stackrel{\$}\gets \sadvbone^{\ms O'_1}(z);b\stackrel{\$}\gets \{0,1\};\\
&\qquad c^*\stackrel{\$}\gets \edem(1^\lambda,k,m_b);\sadvbtwo^{\ms O'_2}(c^*,st)=b].\end{align*}
}
We bound $Adv^{ind\text{-}q_e\text{-}cea}_{\ike,\advb}(\lambda)$ using the defined experiments. Since the iKEM's key is $\sigma(\lambda)$-IND-$q_e$-CCA secure we have:
\begin{align}\nonumber
    |\pr[T_1]-\pr[T_0]|&\leq Adv^{kind\text{-}q_e\text{-}cea}_{\ikem,\advb}(\lambda)\\\label{eq:7}
    &\leq Adv^{kind\text{-}q_e\text{-}cca}_{\ikem,\advu}(\lambda)\leq \sigma(\lambda),
\end{align}

where $\advu$ is a computationally unbounded adversary and by definition is at least as powerful as the computationally bounded adversary $\advb$ (i.e., $Adv^{kind\text{-}q_e\text{-}cea}_{\ikem,\advb}(\lambda)\leq Adv^{kind\text{-}q_e\text{-}cea}_{\ikem,\advu}(\lambda)$).
Using the triangular inequality on (\ref{eq:7}) we have:
\begin{align}\nonumber
    \pr[T_0]-\pr[T_1]&\leq|\pr[T_0]-\pr[T_1]|\leq \sigma(\lambda)\\\label{eq:7'}
    &\Rightarrow \pr[T_0]\leq  \pr[T_1]+\sigma(\lambda),
\end{align}
and from Definition~\ref{def:demsec}, we have: 
\begin{align}\nonumber
&\pr[T_1]\leq Adv^{ind\text{-}ot}_{\dem,\advb}(\lambda)+\frac{1}{2}\leq\sigma'(\lambda)+\frac{1}{2}\\\label{eq:8}
&\stackrel{(\ref{eq:7'})}\Rightarrow \pr[T_0]\leq \sigma(\lambda)+\sigma'(\lambda)+\frac{1}{2} 
\end{align}
Note that for each encryption query in EXP-1 a new key is sampled and therefore $\pr[T_0]$ is bounded by one-time  advantage of the DEM scheme.

According to Definition~\ref{def:she}, $Adv^{ind\text{-}q_e\text{-}cca}_{\ike,\advb}(\lambda)=|\pr[T_0]-\frac{1}{2}|$. By using (\ref{eq:8}) we have:
\begin{equation}
    Adv^{ind\text{-}q_e\text{-}cea}_{\ike,\advb}(\lambda)\leq\sigma(\lambda)+\sigma'(\lambda)\ \ \qed
\end{equation}

\section{A   Construction of iKEM }\label{sec:const}

  OW-SKA  with security definition given as  Definition~\ref{def:owska} is an  IND-OT secure iKEM and so using Corollary~\ref{cor}, results in an IND-OT secure HE.  This is the weakest security notion for encryption systems. Stronger security will be when iKEM security is against an adversary with access to the encryption oracle  (i.e. $q_e$-CEA). 
In this section, we  build on an existing construction of OW-SKA \cite{sharif2020}  with security satisfying Definition~\ref{def:owskasec}, to construct an iKEM with IND-$q_e$-CEA security.
The protocol  analysis provided a lower bound on the key length (finite length analysis).  As shown below, providing  an IND-$q_e$-CEA security for an  iKEM that is based on this protocol requires longer initialization string,

The iKEM construction below is based on the OW-SKA constriction in \cite{sharif2020}.  

\begin{construction}\label{construction}
{The iKEM $\ikem_\mathsf{OWSKA}$.}
   The iKEM $\ikem_\mathsf{OWSKA} = (\ikemg, \ikeme, \ikemd)$ %
   is defined as follows:

Suppose  ${P}_{XYZ}$ is the distribution that is used to generate correlated samples $X,Y$ and $Z$), and let $\{\ms{h}_s:\mc X\to \{0,1\}^t\}_{s\in\mc S}$ and $\{\ms{h'}_{s'}:\mc X\to \{0,1\}^\ell\}_{s'\in\mc S'}$  be  two strong universal hash families (UHFs).
    Also let 
    $\mc C=\{0,1\}^t\times \mc S\times \mc S'$ and $\mc K=\{0,1\}^{\ell}$ denote the sets of ciphertexts and keys, respectively\footnote{We note that $t$ and $\ell$ are both functions of $\lambda$ and parametrized by ${P}_{XYZ}$, the input probability distribution.}.
    The relation between $t,\ell$,  and correctness and security parameters of the iKEM
  is given in  Theorems~\ref{theo:reliability} and \ref{theo:cpa}). 
  These theorems adapt and modify {Theorem 2  in \cite {sharif2020} for iKEM security model.}
   
   {Let   $\{\ms{h}_s:\mc X\to \{0,1\}^t\}_{s\in\mc S}$ and $\{\ms{h'}_{s'}:\mc X\to \{0,1\}^\ell\}_{s'\in\mc S'}$  be  two strong universal hash families (UHFs).}
 
  The iKEM's three algorithms are as follows.

   \begin{itemize}
    \item  
   $ \ikemgo(1^{\lambda}, {P}_{XYZ})$: 
   {For a distribution  ${P}_{{XYZ}}$, a trusted sampler samples the distribution independently $n$ times, and 
   gives 
   \remove{
    The generation algorithm chooses an appropriate  $\mathrm{P}^{n'}_{{XYZ}}$ from $\mc{P}_{XYZ}=\{\mathrm{P}^{n'}_{{XYZ}}|n'\in \mathbb{N}\}$ according to $\lambda$, and 
    samples the distribution
    to output 
    }
    the triplet vectors $\vect x, \vect y$ and $\vect z$ of correlated  samples,   privately 
     to Alice, Bob and Eve, respectively. 
    That is
    $$(\vect{x,y,z})=( x^n, y^n, z^n) \stackrel{\$}\gets \ikemgo(1^{\lambda},{P}_{XYZ}).$$
  }
    \item $\ikemeo(\vect x)$: The encapsulation algorithm   $\ikeme(\cdot)$  takes as input $ x$,  
    samples $s'\stackrel{\$}\gets \mc S'$ and $s\stackrel{\$}\gets \mc S$ for  the seed of the strongly universal hash functions,
and  generates the key $k={\ms{h'}}_{s'}(\vect x)$  and the ciphertext $c=(\ms{h}_s(\vect x),s',s)$,
   Thus
\[
(c,k)=\big((\ms{h}_s( \vect x),s',s),{\ms{h'}}_{s'}(\vect x)\big)\stackrel{\$}\gets\ikemeo(1^\lambda,\vect x).\]
%
   \item  $\ikemdo(\vect y,c)$: The decapsulation  algorithm $\ikemd(\cdot,\cdot)$ takes the private input of Bob, 
   $\vect y$,  and the ciphertext $(\ms{h}_s(\vect x),s',s)$ as input,  and outputs the key $h_{s'}(\vect x)$ or $\perp$. 
   We have:
 \[
 k=({\ms{h'}}_{s'}( \vect x))\gets\ikemdo\big(1^\lambda,\vect y,(\ms{h}_s(\vect x),s',s)\big). \]
 
    The decapsulation algorithm works as follows:
      \begin{enumerate}
    \item Parses the received ciphertext to $(g,s',s)$, where $g$ is a $t$-bit string. 
    \item  Defines the set,
            \begin{align}
             \mathcal{T}(\vect {X}|\vect{ y}) \triangleq \{\vect x :-\log {P}^{n}_{{X}|{Y}} (\vect x|{\vect y}) \leq \nu\},\label{eq:T_list}
        \end{align} 
        {and for each vector $\hat{\vect x}\in \mc T(\vect X|\vect y)$, checks $g \stackrel{?}=\ms{h}_s(\hat{\vect x})$.  }
        \item {Outputs $\hat{\vect x}$ if there is a unique value  $x$ that  satisfies $g=\ms{h}_s({\hat{\vect x}})$; }Else outputs $\perp$.
     \end{enumerate}
   The value  $\nu$ depends on the correlation of $\vect x$ and $\vect y$ where higher correlation corresponds to smaller  $\nu$,
    and smaller set of candidates  
    (see Theorem~1 for the precise relationship).
   
If successful, the decapsulation algorithm outputs a key $k={\ms{h'}}_{s'}(\hat{\vect x})$;
otherwise it outputs $\perp$. 
\end{itemize}
\end{construction}

Let  $\tilde{H}_\infty(\vect X| \vect Y)$ and $\tilde{H}_\infty(\vect X|\vect Z)$,  denote average conditional min-entropies of  the random variables $\vect{X,Y}$ and $\vect Z$.
Theorem~\ref{theo:reliability}  is based on \cite[Theorem 2]{sharif2020} and gives the minimum length of the ciphertext to bound the  error probability of the protocol by $\eps(\lambda)$, and for a given ciphertext length, gives the maximum number of key bits  that can be established by the $\ikem$ with  adversary's advantage $Adv^{kind}_{\ikem_{\ms{OWSKA}},\msa}$ bounded  by $\sigma(\lambda)$, assuming 
adversary does not have {\em any} 
oracle access (encapsulation or decapsulation). 
{The proof of the theorem skips  the last step of ``entropy smoothing'' \cite{Holenstein2011} that was used in \cite{sharif2020} 
%
for proving capacity achieving results.
}
 
\remove{
Note that in contrast to \cite{sharif2020},  here we are not interested in achieving the secrecy capacity of the setting and for simpler representation, we can
 skip the last round of ``entropy smoothing'' \cite{Holenstein2011} in deriving the relations between the scheme parameters and 
 use min-entropy instead of the Shannon entropy as the measure of correlation between Alice and Bob's random variables. 
}

\begin{theorem} 
\label{theo:reliability} 
{Let  $t$ (the output length of $\ms h_s(.)$ in Construction~\ref{construction}) be chosen to satisfy $t\geq 2\tilde{H}_\infty(\vect X|\vect Y)/\eps(\lambda)-\log\eps(\lambda)-1$. 
Then the iKEM $\ikem_{\ms{OWSKA}}$ establishes a  
secret key of length    $\ell \leq \tilde{H}_\infty(\vect X|\vect Z) - t+2\log \sigma(\lambda)+2$ that 
is $\eps(\lambda)$-correct and $\sigma(\lambda)$-IND-OT secure.
}

the length of the established key using the iKEM $\ikem$, satisfies $\ell \leq \tilde{H}_\infty(\vect X|\vect Z) - t+2\log \sigma(\lambda)+2$. 
\end{theorem}

{$q_e$-CPA Security}  for HE schemes requires  iKEM to be 
secure against a  
$q_e$-CEA attacker.
This is achieved by bounding the leaked information due to the $q_e$ CEA queries, and then use privacy amplification \cite{bennett1988privacy} to remove the leaked information. This results in a shorter established key.

\begin{theorem}\label{theo:cpa}
The iKEM $\ikem_{\ms{OWSKA}}$ establishes a key of 
 length  $\ell \leq \frac{2+2\log\sigma_{e}(\lambda)+\tilde{H}_\infty(\vect X|\vect Z)}{q_{e}+1}-t-\log(q_e/\sigma_e(\lambda))$ that is $2\sigma_e(\lambda)$-indistinguishable from random by an adversary with access to $q_e$ encapsulation queries  ($2\sigma_e(\lambda)$-IND-$q_e$-CEA).

\end{theorem}

\textit{Proof}.
A query to the encapsulation oracle gives 
  a pair  of 
  key and ciphertext 
  $(c,k)$ to the adversary.  Let the vector $\mathbf{v}^{q_e\text{-}cea} =( v_1^{cea}, \cdots, v^{cea}_{q_e})$ be  the 
  adversary's received responses to its queries. 
The remaining uncertainty about $ X$ that can be used for the key 
is 
 $H_\infty(X|\mb{V}^{cea}=\mb{v}^{cea})$. 
Let  $v_i^{cea}=(c_i,k_i)$, be the $i^{th}$ query's response, values of $S$ and $S'$  (in $\ms{h}_S(\vect X)$ and ${\ms{h'}}_{S'}(\vect X)$) in the $i^{th}$ query's response   be 
  $s_i$ and $s'_i$, and $c_i$ be $({c_0}_i,s_i,s'_i)$.
  From Lemma~\ref{lem:entropy}(b), for RVs $C_{0_i}$  and $K_i$  that are distributed over $\{0,1\}^t$ and $\{0,1\}^\ell$, respectively, we have $\tilde{H}_\infty(\vect{X|Z},C_{0_i},K_i)\geq \tilde{H}_\infty(\vect{X|Z})-t-\ell$, and from Lemma~\ref{lem:entropy}(a), $\tilde{H}_\infty(\vect{X|Z},C_{0_i}=c_{0_i},K_i=k_i)\geq \tilde{H}_\infty(\vect{X|Z},C_{0_i},K_i)-log(1/\delta)$, with probability at least $1-\delta$ over the choice of ($c_{0_i},k_i$). Let $\delta=\frac{\sigma_e(\lambda)}{q_e}$. Then, for each query
\begin{align}\label{seq:1}\nonumber
    \tilde{H}_\infty(\vect{X|Z},v_i^{cea})&=\tilde{H}_\infty(\vect{X|Z},C_i=c_{0_i},K_i=k_i)\\
    &\geq \tilde{H}_\infty(\vect{X|Z})-t-\ell-log(q_e/\sigma_e(\lambda)),
\end{align}
with probability at least $1-\frac{\sigma_e(\lambda)}{q_e}$.
This is the adversary's  minimum uncertainty about $\vect X$ after making a single query to the encapsulation  oracle.
The bound of (\ref{seq:1}) shows that after adversary sees the output of a query to the encapsulation oracle, the min-entropy of $\vect{X|Z}$ will be decreased by at most $t+\ell+\log(q_e/\sigma_e(\lambda))$  with probability at least $1-\frac{\sigma_e(\lambda)}{q_e}$.
Here,  probability is { over the randomness of the encapsulation.
That is, with probability at most $\frac{\sigma_e(\lambda)}{q_e}$, there is no guarantee on the amount of leaked information. }This means for some encapsulation oracle output, the min-entropy of $\vect{X|Z}$ can become very small. However, such oracle outputs will happen only with probability $\frac{\sigma_e(\lambda)}{q_e}$.


In Construction~\ref{construction}, the randomness of the encapsulation oracle is from the random choice of seeds, and these seeds are IID (independently sampled random at uniform) to answer each CEA query. Therefore,
  After $q_{e}$ queries we have $\tilde{H}_{\infty}(\vect{X|Z},\mathbf{V}^{q_e\text{-}cea}=\mathbf{v}^{q_e\text{-}cea})\geq \tilde{H}_{\infty}(\vect{X|Z})-q_{e}(t+\ell+\log(q_e/\sigma_e(\lambda)))$ with probability at least $(1-\frac{\sigma_e(\lambda)}{q_e})^{q_e}$, and since from Lemma~\ref{lem:glhl}
\begin{align}\nonumber
  \sd\big(&(Z,\ms{h}_S({ X}),S,S',({\ms{h'}}_{S'}(\vect X));\\\label{eq:sigma-sec}
  &(Z,\ms{h}_S({ \vect X}),S,S',U_\ell)\big)\leq \frac{1}{2}\sqrt{2^{t+\ell-\tilde{H}_\infty(\vect{X|Z})}},
\end{align}
we have:
\begin{align*}
  \sd\bigg(\Big(&\vect Z,S^{q_e+1},S'^{q_e+1},{\ms{h}}_{S}(\vect{X}),\ms h'_{S'}(\vect{X}),\vect{v}^{q_e\text{-}cea}\Big);\\
  &\Big(\vect Z,S^{q_e+1},S'^{q_e+1},\ms h_S(\vect{X}),U_\ell,\vect{v}^{q_e\text{-}cea}\Big)\bigg)\\
  &\leq \frac{1}{2}\sqrt{2^{(q_e+1)(t+\ell+\log(q_e/\sigma_e(\lambda)))-\tilde{H}_\infty(\vect{X|Z})}}, 
\end{align*}
with probability $1-\frac{\sigma_e(\lambda)}{q_e})^{q_e}$. Since
 $\ell \leq \frac{2+2\log\sigma_{e}(\lambda)+\tilde{H}_\infty(\vect{X| Z})}{q_{e}+1}-t-\log(q_e/\sigma_e(\lambda))$, the above statistical distance is bounded by $\sigma_e(\lambda)$ with probability $(1-\frac{\sigma_e(\lambda)}{q_e})^{q_e}$and  by 1 otherwise. Thus we have:
\small\begin{align*}
  &\sd\bigg(\Big(\vect Z,S,S',{\ms{h}}_{S}(\vect{X}),\ms h'_{S'}(\vect{X}),\mathbf{V}^{q_e\text{-}cea}_{\msa}\Big);\\
  &\Big(Z,S,S',\ms h_S(\vect{X}),U_\ell,\mathbf{V}^{q_e\text{-}cea}_{\msa}\Big)\bigg)\leq (1-\frac{\sigma_e(\lambda)}{q_e})^{q_e}\sigma_{e}(\lambda)+\\
  &\quad\qquad\qquad\qquad\qquad\qquad\qquad\qquad\big(1-(1-\frac{\sigma_e(\lambda)}{q_e})^{q_e}\big)\\
  &\qquad\qquad\qquad\qquad\qquad\qquad\qquad\stackrel{(1)}\leq\sigma_{e}(\lambda)+\sigma_e(\lambda)\leq 2\sigma_e(\lambda),
\end{align*}
\normalsize
where $(1)$ inequality is since $1-\frac{\sigma_e(\lambda)}{q_e}\leq 1$ and $1-(1-\frac{\sigma_e(\lambda)}{q_e})^{q_e}\leq \sigma_e(\lambda)$ due to Bernoulli's inequality stating  for $t\geq 1$ and $0\leq x\leq 1$, inequality $xt\geq 1-(1-x)^t$ holds. Finally, for $C^*=(h_S(\vect{X}),S',S)$, the inequality (\ref{eq:ikemsd}) is satisfied .
That is  we have $2\sigma_{e}(\lambda)$-indistinguishability against $q_e$ CEA.$\ \blacksquare$

\section{Concluding Remarks}\label{sec:conclude}

We initiated the study of hybrid encryption in preprocessing
model, defined its security and proved a composition theorem for iKEM and DEM to achieve $q_e$-CPA security. Our work provides opportunities for constructing post-quantum secure hybrid encryption systems that are do not rely on computational assumptions.
A secure DEM that can be   constructed using existing secure block cipher  algorithms, and replaces the public-key part with an iKEM with information-theoretic security.

\textbf{On the composability of iKEM:}
Informally, the correctness and security conditions of an iKEM 
can be rephrased in a composablity framework such as the UC framework \cite{canetti2001universally} or the Constructive Cryptography framework \cite{maurer2011constructive} as follows: 
In the ideal world   iKEM constructs a shared secret key  using resources (correlated randomness  in the preprocessing phase), and in the real world the key is established  using an authenticated communication channel, such that the two worlds are indistinguishable  for a computationally unbounded environment. 


This can be seen  by combining  correctness and indistinguishability\footnote{We show this for an IND-OT iKEM. The proof for IND-CEA security is the same.} conditions using a single  bound on the statistical distance of the random variables in the  two worlds, as shown below. 
\begin{lemma}
For an $\eps(\lambda)$-correct, $\sigma(\lambda)$-IND-OT iKEM in Definition~\ref{def:ikem} with security parameter $\lambda$, input probability distribution $\mc P$, and key space  $\{0,1\}^{\likem_{\mc P}(\lambda)}$,  let $k_A$ and $k_B$ be the keys obtained by Alice and Bob, and $K_A$ and $K_B$ be the corresponding random variables, respectively. {That is,
$\ikeme(x).key =k_A$} and $\ikemd(y,c)=k_B$. Then
\begin{align}\nonumber
&\sd\big((Z,C^*,K_A,K_B);(Z,C^*,U_{\likem(\lambda)},U_{\likem_{\mc P}(\lambda)})\big)\\
&\qquad\qquad\qquad\qquad\qquad\qquad\qquad\qquad\leq \eps(\lambda)+ \sigma(\lambda),\label{eq:uc}
\end{align}
\end{lemma}
\textit{Proof.} According to the correctness condition, $K_A$ and $K_B$ are identical unless an error accrues with probability at most $\eps(\lambda)$. Therefore, by the application of \cite[Lemma 6.2]{cramer2003design} we have:
\begin{equation}\label{eq:ucp1}
\sd\big((Z,C^*,K_A,K_A);(Z,C^*,K_A,K_B)\big)\leq \eps(\lambda).
\end{equation}
On the other hand, a $\sigma(\lambda)$-IND-OT iKEM satisfies:
\begin{align}\nonumber
&\sd\big((Z,C^*,K_A);(Z,C^*,U_{\likem_{\mc P}(\lambda)})\big)\leq \sigma(\lambda)\stackrel{(\ref{eq:sd})}\Rightarrow\\\nonumber
& \sd\big((Z,C^*,K_A,K_A);(Z,C^*,U_{\likem_{\mc P}(\lambda)},U_{\likem_{\mc P}(\lambda)})\big)\\\label{eq:ucp}
&\quad\qquad\qquad\qquad\qquad\qquad\qquad\qquad\qquad\qquad \leq \sigma(\lambda).
\end{align}

Now from the triangular inequality
\begin{align}\nonumber
&\sd\big((Z,C^*,K_A,K_B);(Z,C^*,U_{\likem(\lambda)},U_{\likem_{\mc P}(\lambda)})\big)\\\nonumber
&\leq\sd\big((Z,C^*,K_A,K_B);(Z,C^*,K_A,K_A)\big)\\
&\quad+\sd\big((Z,C^*,K_A,K_A);(Z,C^*,U_{\likem(\lambda)},U_{\likem_{\mc P}(\lambda)})\big)\label{eq:tr}
\end{align}
The first term of the above inequality is bounded by (\ref{eq:ucp1}) and the second term is bounded by (\ref{eq:ucp}). Thus, we have (\ref{eq:uc}).
$\blacksquare$

\noindent
\textbf{Computationally unbounded HE:}
A natural extension of our work is considering hybrid encryption in preprocessing model when DEM has information-theoretic security.
Shannon's one-time pad (OTP)  
 is the only \textit{deterministic} symmetric encryption scheme that is secure against a computationally unbounded adversary and can be used as DEM in such an HE system. The combined system effectively allows a key to be established, and used with an OTP to provide
 $\sigma(\lambda)$-IND-OT security.
 Extending our work to include probabilistic DEM is an interesting future direction that will allow  probabilistic information-theoretic symmetric key encryption schemes such as  \cite{dodis2005entropic,fool2006,sharifian2019modular} 
 to  be used for DEM.

\bibliographystyle{IEEEtran}

\IEEEtriggercmd{\enlargethispage{-10cm}}

\end{document}